\documentclass[reprint,amsmath,amssymb,aps,floatfix]{revtex4-2}
\usepackage{graphicx}
\usepackage{dcolumn}
\usepackage{bm}
\usepackage{hyperref}
\usepackage{xcolor}
\usepackage{caption}
\usepackage{subcaption}
\usepackage{multirow}
\usepackage{tikz}

\newcommand{\sublabel}[2]{%
  \begin{tabular}[t]{@{}l@{}}
    {\bfseries (#1)}\\
    #2
  \end{tabular}%
}

\begin{document}

\title{Crystal Fractional Graph Neural Network for Energy Prediction of High-Entropy Alloys}
\author{Takanori Kotama}
\affiliation{
    School of Informatics, Nagoya University, Nagoya 4648601, Japan
}
\author{Yang Huang}
\affiliation{
  University of Science and Technology of China, Hefei 230026, China
}
\affiliation{
  Suzhou Institute for Advanced Research, University of Science and Technology of China, Suzhou 215213, China
}

\begin{abstract}
High-entropy alloys (HEAs) have attracted growing attention for their exceptional mechanical and thermal properties arising from complex atomic configurations.
In this paper, we propose crystal fractional graph neural network for predicting the energy of high-entropy alloys by explicitly integrating both local atomic environments and global compositional information.
The model consists of three components: a crystal graph neural network, which employs graph attention network layers to learn local interactions among 16 on-site atoms within the crystal lattice; fractional neural network, a fully connected network that embeds the global fraction of constituent elements; and feature fusion neural network, which fuses the outputs of the two submodels to predict the total crystal energy.
We train the model on a dataset of 1,049 crystal structures and validate it on 198 quaternary structures, optimizing all hyperparameters via Optuna.
Our results show that our model achieves an RMSE comparable to first-principles calculations and maintains high accuracy even for low-energy configurations.
However, the model exhibits limitations in handling large crystal cells, which we aim to address in future work to extend its applicability to more complex systems.
\end{abstract}

\maketitle

\section{Introduction}

High-entropy alloys (HEAs) are advanced materials notable for their exceptional mechanical strength and hardness, and also exhibit impressive thermal stability, maintaining their properties at high temperatures, and enhanced corrosion resistance~\cite{Yeh2004,Cantor2004,Senkov2011,Tsai2014,Ye2016,Lu2016,George2019}.
Additionally, they are highly resistant to radiation-induced damage, making them suitable for challenging environments like nuclear reactors and space applications~\cite{El-Atwani2019,Xu2021,Dada2019}.
Their durability in abrasive conditions further enhances their versatility and reliability in extreme settings.
Despite these promising attributes, research on HEAs faces significant challenges due to their vast compositional space.
The wide variety of possible element combinations makes it difficult to systematically explore and optimize these alloys.
This complexity necessitates advanced computational and experimental techniques to effectively navigate and understand the expansive landscape of HEA compositions.

Experimental investigations of HEAs can be prohibitively expensive and time-consuming, often requiring days or even weeks to yield results~\cite{Dabrowa2016,Ondicho2019}.
This extended timeframe and high cost can significantly slow down the discovery and development of new HEAs.
To address these challenges, computational methods have become invaluable tools in the exploration of HEAs.
Techniques such as first-principles energy calculations and molecular dynamics simulations enable researchers to predict and analyze the properties of these alloys more efficiently.
First-principles calculations provide detailed insights into the fundamental interactions at the atomic level, while molecular dynamics simulations allow for the investigation of material behavior over time and under various conditions.
By leveraging these computational approaches, researchers can accelerate the discovery process, optimize alloy compositions, and reduce the reliance on costly and lengthy experimental procedures.

With the advent of machine learning methods, research into HEAs has entered a new era.
These techniques have become integral to various aspects of HEA research, greatly enhancing both the speed and scope of investigations.
Machine learning is now pivotal in modeling interatomic potentials for HEAs, enabling accurate simulations of atomic interactions and material properties~\cite{Wu2024,Ladygin2020,Byggm2021,Byggm2022}.
This allows researchers to predict physical attributes like mechanical strength~\cite{Wang2023} more efficiently.
Furthermore, machine learning aids in studying point defects~\cite{Wang2022,Byggm2021} and dislocations~\cite{Wang2024,Wu2024} within HEAs, providing insights into their effects on material performance.

Historically, HEA studies have relied heavily on on-site potential models, such as the cluster expansion method~\cite{kikuchi1951,sanchez1984,deFontaine1989}, and effective pair interaction model~\cite{liu2019,Zhang2020}.
These models were designed to handle substitutional systems by representing interactions between different alloying elements and predicting their effects on the material's properties, such as order-disorder phase transitions~\cite{Kadkhodaei2021,Kim2023}.
These potential models, however, fall short of first-principle accuracy and are significantly constrained by the limitations of their compositional space.
Typically, these models rely on fixed composition datasets, which restrict their ability to generalize across a wider range of compositions~\cite{Kim2023,liu2019}.

To overcome these limitations, we propose a fast and accurate on-site machine-learning model, Crystal Fractional Graph Neural Network (CrysFracGNN), designed specifically for HEAs and covering the entire compositional space.
The model consists of three components: CrystalGNN, which uses graph attention network (GAT) layers to capture local atomic interactions within a 16-atom BCC supercell; FractionFNN, a fully connected network that encodes the global elemental fractions; and ConcatFNN, which combines both representations to predict total crystal energy.
Our approach seeks to integrate the precision of advanced machine learning techniques with the simplicity and efficiency of traditional on-site models, enabling more effective and manageable simulations of HEA systems.

This paper is organized as follows.
Section~\ref{sec:methodology} describes the dataset, first-principles calculations, and the CrysFracGNN architecture and hyperparameter optimization.
Section~\ref{sec:results} presents the prediction results for standard 16-atom structures and large-scale supercells.
Section~\ref{sec:discussion} discusses model performance and limitations and concludes with future directions.

\section{Methodology}
\label{sec:methodology}

\begin{figure}[htpb]
  \centering
  \includegraphics[width=0.9\linewidth]{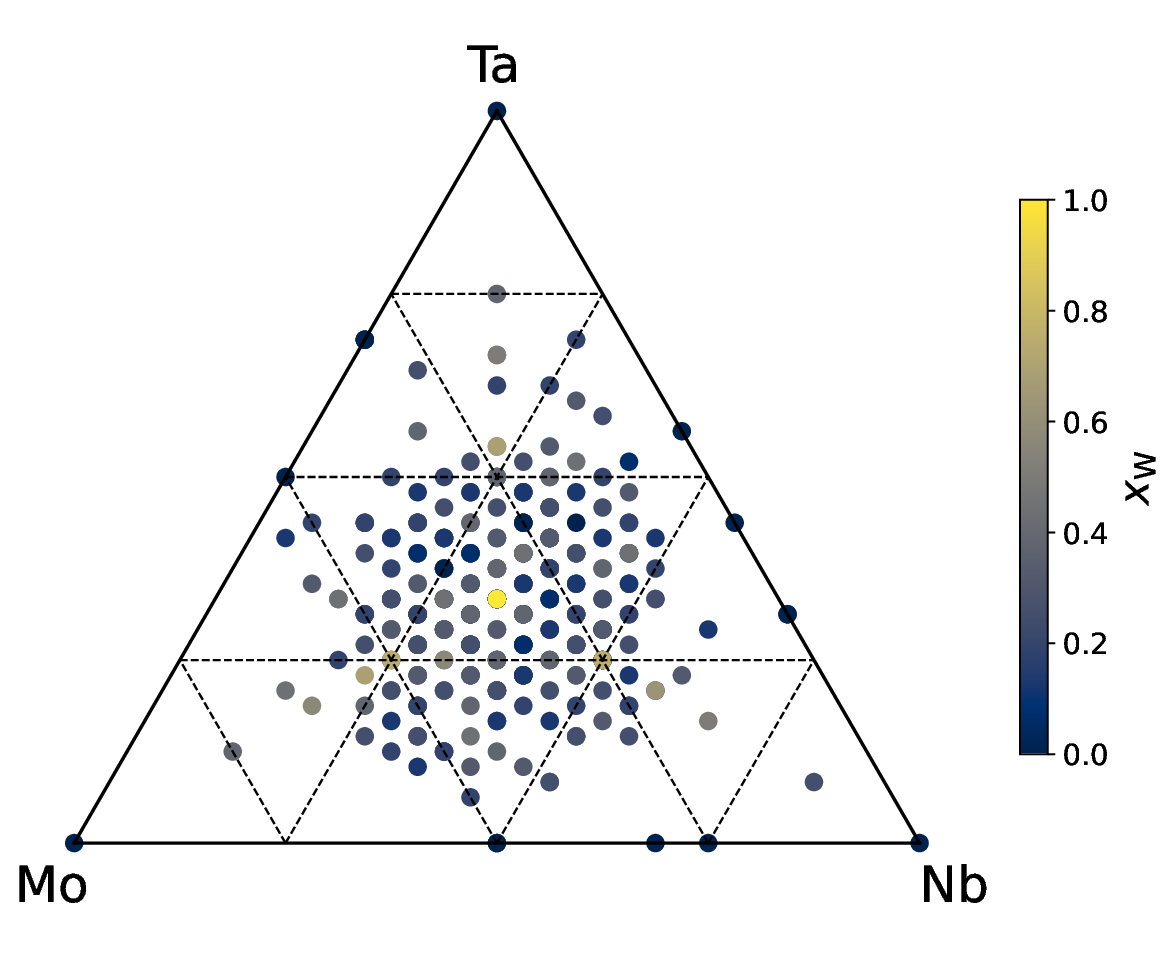}
  \caption{A distribution of data points of Mo-Nb-Ta-W.}
  \label{fig:distribution}
\end{figure}

In particular, our model is examined in the Mo-Nb-Ta-W HEA system, where the Nb-Mo-Ta-W HEA exhibits a body-centered cubic (BCC) structure.
We have created a dataset that includes 598 quaternary crystal structures, 16 ternary structures, 24 binary structures, and 4 single-element structures.
Each entry in the dataset provides comprehensive details on the types and positions of atoms within the crystal lattice.
The explanatory variables describe the types of atoms present and their spatial arrangement, while the objective variable represents the total energy of the crystal.
All structures were generated by randomly varying the composition and atomic decoration within the supercell, which contains 16 atoms arranged in an ideal B2 configuration as shown in Fig.~\ref{fig:scell}~(a).
The overall distribution of the composition of our dataset can be seen in Fig.~\ref{fig:distribution}.

\subsection{Datasets}

\begin{figure}[htpb]
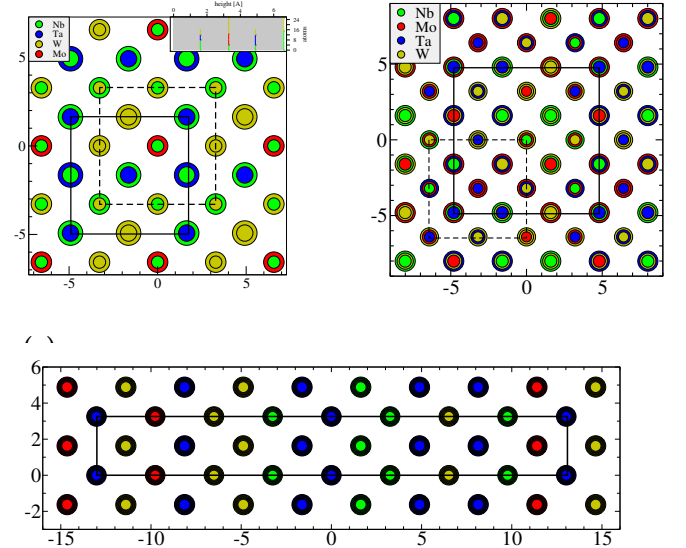

  \centering
  \sublabel{a}{\includegraphics[width=.45\linewidth]{figs/xyz.eps}}\hfill
  \sublabel{b}{\includegraphics[width=.45\linewidth]{figs/54-atom.eps}}\\
  \sublabel{c}{\includegraphics[width=.915\linewidth]{figs/1024-atom_2.eps}}
  \caption{Atomic decoration of (a) a 16-atom structure, (b) a 54-atom structure, and (c) a $8\times 1\times 1$ supercell.}
  \label{fig:scell}
\end{figure}

We use 400 quaternary crystal structure datasets, 16 ternary crystal structure datasets, 24 binary crystal structure datasets, and 4 single-element crystal structure datasets for training data and 198 quaternary crystal structure datasets for validation data.
To be able to predict low-energy crystal structures, we add 605-point low-energy datasets to the training data.
We use a total of 1049 crystal structure datasets for training data and 198 quaternary crystal structure datasets for validation data.
For evaluating the performance of the model under symmetry operations, we created 435456-point datasets for test data using physical rules such as rotation invariance and mirror symmetry from 598-point quaternary datasets.
To assess the scalability of our model, we create larger supercells and prepare two distinct datasets: a 54-atom dataset and a 1024-atom dataset.
The 54-atom dataset is generated by constructing a $3\times 3\times 3$ supercell of the B2 unit cell (Fig.~\ref{fig:scell}~(b)), incorporating random compositions and atomic decorations.
For the 1024-atom dataset, we first create a $1\times 1\times 8$ supercell of the B2 unit cell and perform the first-principles calculations, as illustrated in Fig.~\ref{fig:scell}~(c).
Subsequently, the resulting structure is upscaled in the $xy$-direction to achieve the 1024-atom configuration.

\subsection{First-principles calculations}

The initial atomic positions within the cubic cell are set to a standard ideal arrangement to establish a baseline structure.
To estimate the initial densities of these structures, we approximate them based on their corresponding atomic volume,
\begin{equation}
  \label{eq:density}
  \rho = \frac{1}{v_{\rm tot}} = \frac{1}{x_{\rm Mo}v_{\rm Mo} + x_{\rm Nb}v_{\rm Nb} + x_{\rm Ta}v_{\rm Ta} + x_{\rm W}v_{\rm W}},
\end{equation}
where $v_{\rm Mo},v_{\rm Nb},v_{\rm Ta},v_{\rm W}$ are the fitted atomic volumes, respectively, with values of $14.99$, $18.94$, $18.53$, and $14.77$~\AA$^3/$atom.
Full relaxation (lattice parameters and atomic coordinates) is carried out using density functional theory (DFT)~\cite{Kohn1965} as implemented in the Quantum ESPRESSO package~\cite{QE-2009,QE-2017}.
For the calculations, we employed augmented plane wave potentials~\cite{Kresse1999} within the Perdew-Burke-Ernzerhof (PBE) generalized gradient approximation~\cite{Perdew1996} for the exchange-correlation functional, as provided by the PSlibrary~\cite{DALCORSO2014}.
The plane-wave basis energy cutoff and charge density cutoff are set to 80 Ry and 320 Ry, respectively.
A high electronic $k$-point density is used with a $\Gamma$-centered Monkhorst-Pack scheme~\cite{Monkhorst1976}.
For the 16-atom structures, a $4\times 4\times 4$ $k$-point grid is used, while a $3\times 3\times 3$ $k$-point grid is employed for the 54-atom structures.
The energies of all structures are converged to within $0.1$ meV, and the forces are converged to within $1$ meV/\AA.

\subsection{CrysFracGNN Architecture}

\begin{figure*}[htpb]
  \centering
  \includegraphics[width=.7\linewidth]{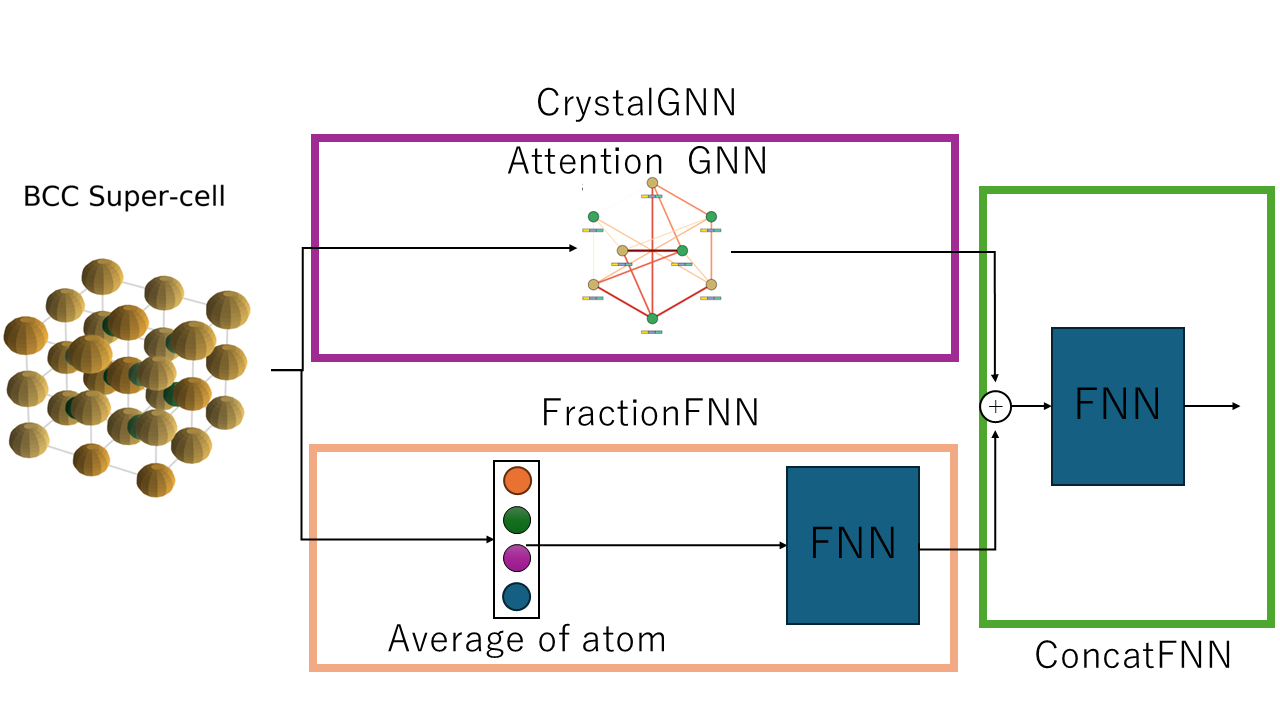}
  \caption{Schematic architecture of CrysFracGNN, composed of CrystalGNN (GAT-based local interaction module), FractionFNN (elemental fraction encoder), and ConcatFNN (energy predictor).}
  \label{fig:architecture_overview}
\end{figure*}

As shown in Fig.~\ref{fig:architecture_overview}, CrysFracGNN is a graph neural network (GNN)-based model tailored to crystal structure data.
GNNs are well suited to this task because they can explicitly represent the interatomic relationships within a crystal lattice.
Xie et al.~\cite{Xie_2018} demonstrated that GNN-based models can accurately predict both the formation energy and the absolute energy of a crystal structure, indicating that such models capture the essential information required for energy prediction.
Building on this line of work, we construct $\text{CrysFracGNN}$, which combines GNN layers with fully connected layers to predict the energy of HEA crystals.
The model consists of three components: $\text{CrystalGNN}$, a GNN-based module that preprocesses the 16-node graph representation of the crystal; $\text{FractionFNN}$, a fully connected network that embeds the global compositional information; and $\text{ConcatFNN}$, a fully connected network that fuses the outputs of the two submodules to produce the final energy prediction.

\begin{figure}[htpb]
  \centering
  \includegraphics[width=.7\linewidth]{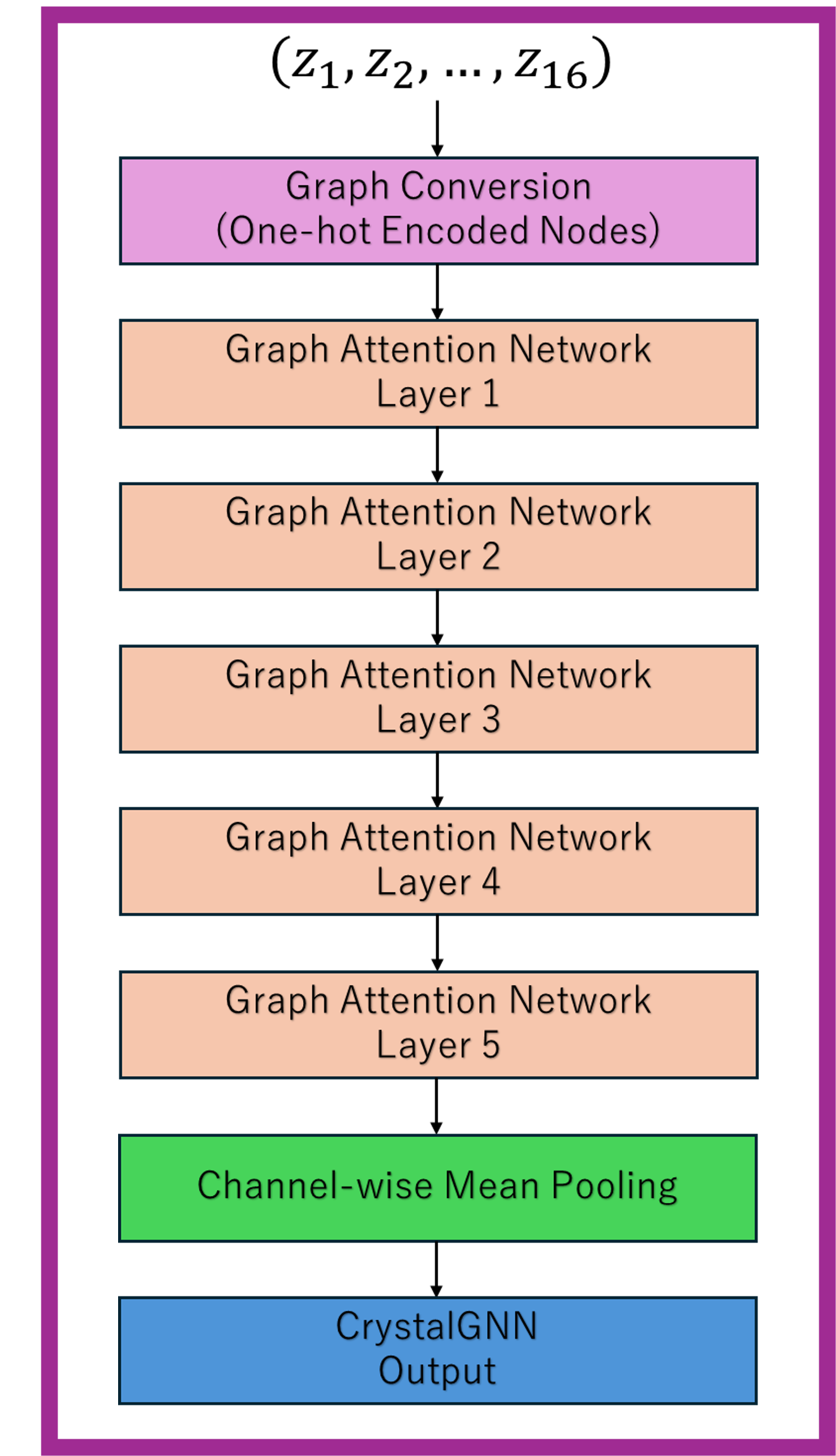}
  \caption{Detailed architecture of the CrystalGNN, which processes atomic graph representations of a 16-atom BCC supercell using five Graph Attention Network (GAT) layers and channel-wise mean pooling.}
  \label{fig:GNN_network}
\end{figure}

\begin{figure}[htpb]
  \centering
  \includegraphics[width=.7\linewidth]{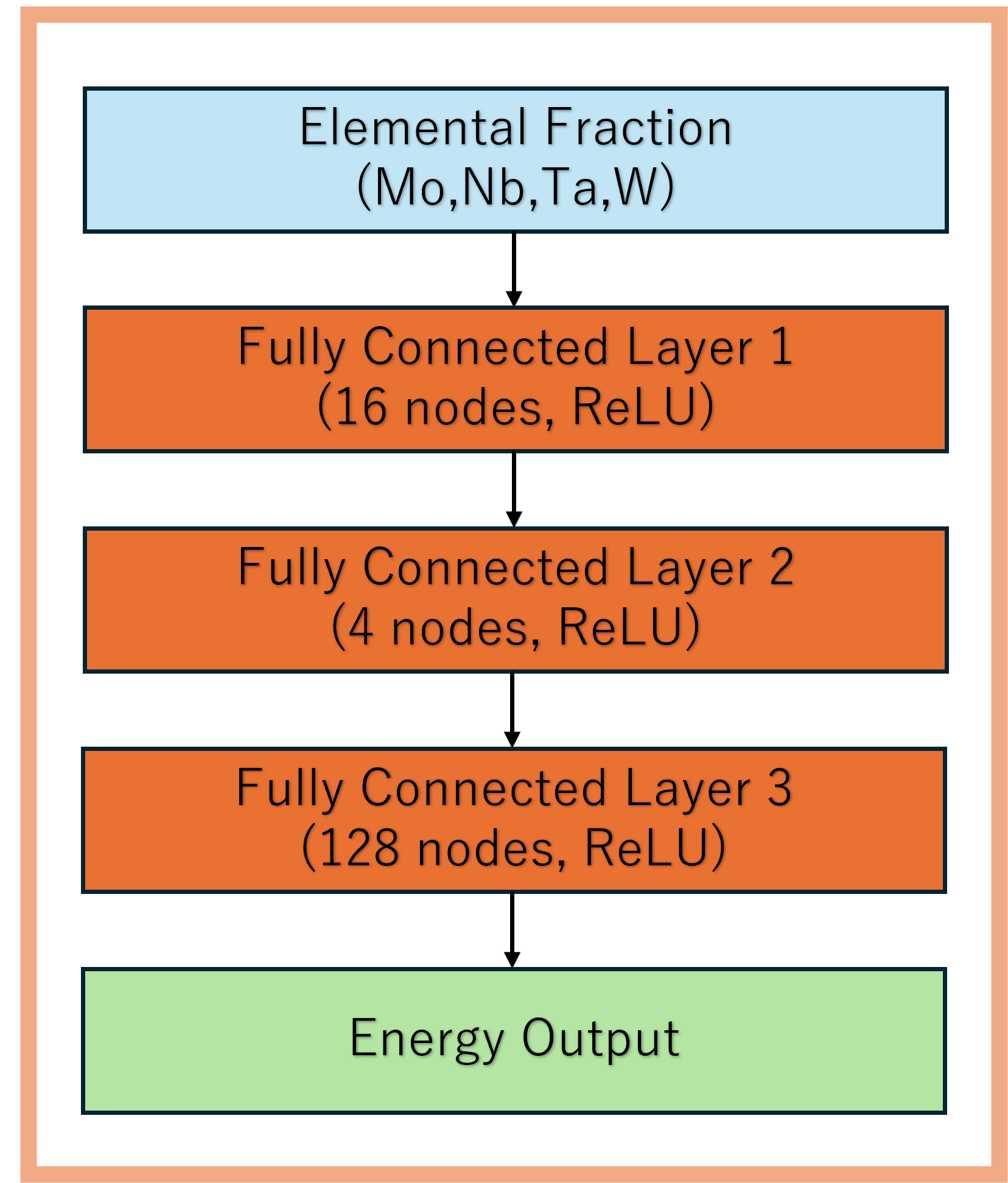}
  \caption{Detailed architecture of FractionFNN, a feed-forward network that encodes the elemental fractions of Mo, Nb, Ta, and W into a compositional embedding.}
  \label{fig:FNN_detail}
\end{figure}

\begin{figure}[htpb]
  \centering
  \includegraphics[width=.7\linewidth]{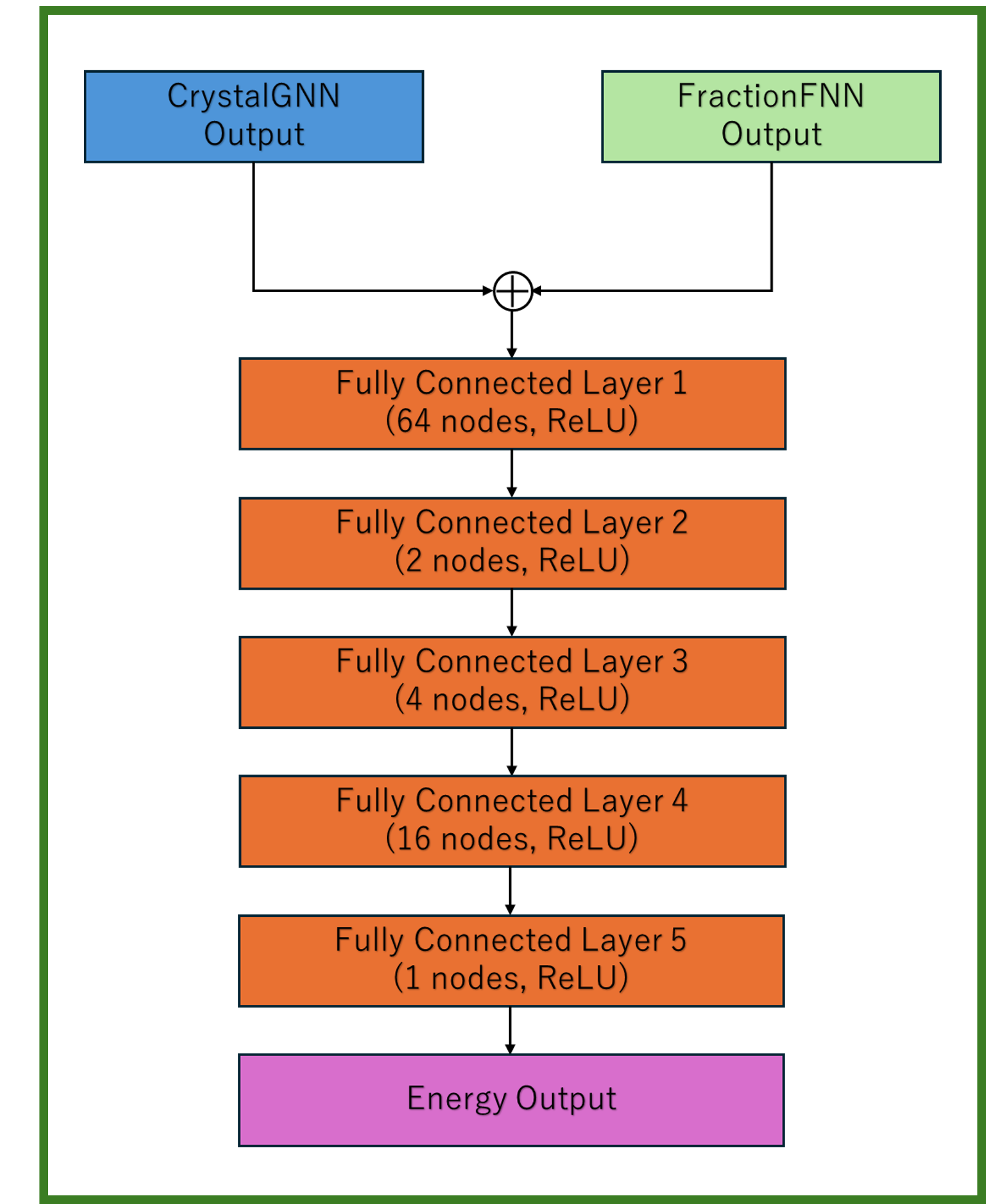}
  \caption{Detailed architecture of ConcatFNN, a fully connected network that predicts the total crystal energy from the concatenated outputs of CrystalGNN and FractionFNN.}
  \label{fig:CrysFracGNN_overview}
\end{figure}

\subsubsection{CrystalGNN Architecture}

$\text{CrystalGNN}$ has several Graph Attention Network (GAT) layers and each layer has a ReLU activation function.
The GAT layer proposed by Veli\v{c}kovi\'{c} et al. makes the GNN layer have different weights for each node~\cite{veličković2018graphattentionnetworks}.
For applying GNN layers, we convert 16 features to graph data which has 16 nodes and input this data to the GNN model.
Using the crystal feature in which the lattice is periodically repeated, we can identify which atoms are nearest to the target atom beyond the unit cell.
If the target atom is in the unit cell, the feature of the target atom is the same as the feature of the target atom which is in the corresponding position in the unit cell.
Therefore, all of the nodes have 8 nodes which are connected to the target node in the unit cell.

Figure~\ref{fig:GNN_network} shows the network architecture for input and output data network for $\text{CrystalGNN}$.
This network has different weights for each node because the weight of the edge expresses the distance between the target atom and the connected atom.
We use two radii for the edge weight, 1 and 2.
The edge weight is 1 if the distance between the target atom and the connected atom is under 0.45~\AA\ and the edge weight is 2 if the distance is over 0.45~\AA\ and under 0.55~\AA.
Applying to 54-atom and 1024-atom structures, we use different radii for the edge weight to preserve the identity of the network.
The edge weight is 1 if the distance between the two atoms is under 0.3~\AA\ and the edge weight is 2 if the distance is over 0.3~\AA\ and under 0.366~\AA\ in the 54-atom structure, and the edge weight is 1 if the distance is under 0.1125~\AA\ and the edge weight is 2 if the distance is over 0.1125~\AA\ and under 0.1375~\AA\ in the 1024-atom structure, respectively.
This model is used for embedding the crystal structure information and atom information to the $\text{ConcatFNN}$.

\subsubsection{FractionFNN Architecture}

As shown in Fig.~\ref{fig:FNN_detail}, $\text{FractionFNN}$ is a fully connected neural network consisting of several layers, each followed by a ReLU activation function.
The input to this network is the four-dimensional composition vector $(x_{\rm Mo}, x_{\rm Nb}, x_{\rm Ta}, x_{\rm W})$, where each component is in the range $[0, 1]$ and is computed as the fractional occurrence of the corresponding element within the crystal structure.
This submodule produces a compositional embedding that is passed to $\text{ConcatFNN}$, providing the global chemical context that complements the local structural information extracted by $\text{CrystalGNN}$.

\subsubsection{ConcatFNN Architecture}

As illustrated in Fig.~\ref{fig:CrysFracGNN_overview}, $\text{ConcatFNN}$ is a fully connected neural network consisting of several layers, each followed by a ReLU activation function.
It takes two inputs: the output of $\text{CrystalGNN}$ and the output of $\text{FractionFNN}$.
To convert the $\text{CrystalGNN}$ output, which has node-wise features, into a fixed-size vector suitable for $\text{ConcatFNN}$, we apply channel-wise mean pooling across the 16 nodes.
This pooling operation aggregates the local structural and atomic features into a single representation for each channel.
As a result, the input dimension of $\text{ConcatFNN}$ equals the sum of the number of channels in the final layer of $\text{CrystalGNN}$ and the number of neurons in the final layer of $\text{FractionFNN}$.
The network then combines these two embeddings to predict the total energy of the crystal structure.

\subsection{Hyperparameter Optimization}

We use Optuna~\cite{akiba2019optuna} for hyperparameter optimization with 100 trials.
Using Optuna, we decide the number of channels and layers that $\text{CrystalGNN}$ has and the number of neurons and layers for $\text{FractionFNN}$ and $\text{ConcatFNN}$.
The range of the number of layers for $\text{CrystalGNN}$ is 0 to 6 and the range of the number of channels is 1 to 256.
The number of channels must be a power of 2.
The range of the number of layers for $\text{FractionFNN}$ and $\text{ConcatFNN}$ is 0 to 6 and the range of the number of neurons is 1 to 256.
The number of neurons must be a power of 2.
The model is trained with the Adam optimizer.
The learning rate is optimized in the range of $10^{-5}$ to $10^{-1}$ and the batch size is 32.
To facilitate model fitting, we use a scheduler which changes the learning rate when the validation loss is not improved for specific epochs.
The specific epochs are optimized in the range of 1 to 100 and the rate of changing the learning rate is optimized in the range of 0.1 to 1.0.
To prevent overfitting, we use early stopping with a patience of 10 and choose the best model that has the lowest validation loss, as we listed in Tab.~\ref{table:model_architecture}.
The loss function we set is MSE and we optimize the model to minimize the loss function.

\section{Results}
\label{sec:results}

\subsection{Use 1247-point datasets}

\begin{table*}
  \centering
  \caption{Summary of CrysFracGNN model architecture and optimizer hyperparameters. The architecture section shows the number of channels (CrystalGNN) or nodes (FractionFNN, ConcatFNN) per layer. The optimizer section shows the learning rate, scheduler patience, and gamma used during training.}
  \label{table:model_architecture}
  \begin{tabular}{ll|ccccc}
    \hline
    \multicolumn{7}{c}{Architecture} \\
    \hline
    Module & Parameter & Layer 1 & Layer 2 & Layer 3 & Layer 4 & Layer 5 \\
    \hline
    CrystalGNN  & channels & 128 & 2   & 256 & 64 & 256 \\
    FractionFNN & nodes    & 16  & 4   & 128 & -- & --  \\
    ConcatFNN   & nodes    & 64  & 2   & 4   & 16 & 1   \\
    \hline
    \multicolumn{7}{c}{Optimizer Hyperparameters} \\
    \hline
    \multicolumn{3}{l|}{Learning Rate} & \multicolumn{2}{c|}{Patience} & \multicolumn{2}{c}{Gamma} \\
    \hline
    \multicolumn{3}{l|}{0.00388} & \multicolumn{2}{c|}{46} & \multicolumn{2}{c}{0.348} \\
    \hline
  \end{tabular}
\end{table*}

Using 1247-point datasets, we optimized the hyperparameters of the model using Optuna.
Table~\ref{table:model_architecture} shows the model architecture and optimizer hyperparameters of CrysFracGNN.

\begin{figure*}
  \centering
  \sublabel{a}{\includegraphics[width=.3\linewidth]{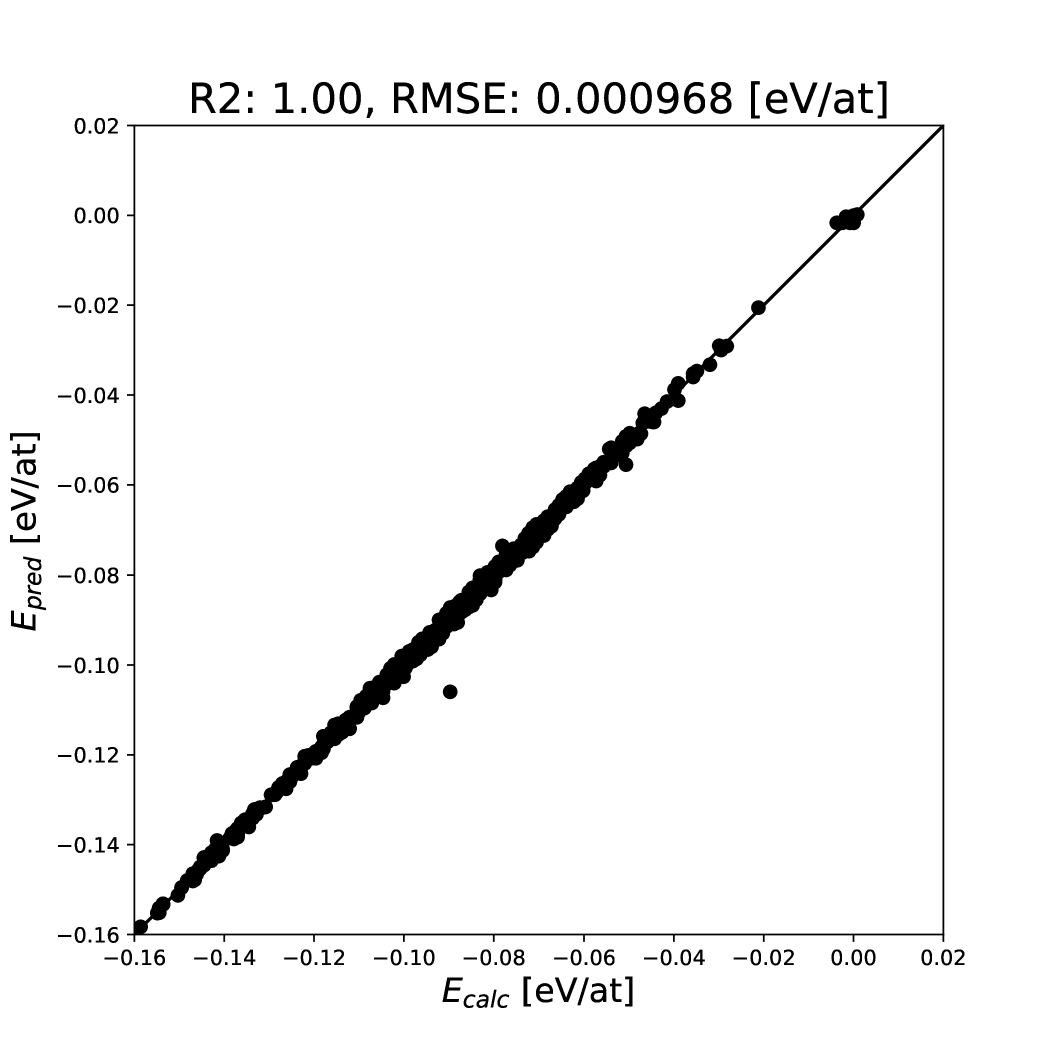}}\hfill
  \sublabel{b}{\includegraphics[width=.3\linewidth]{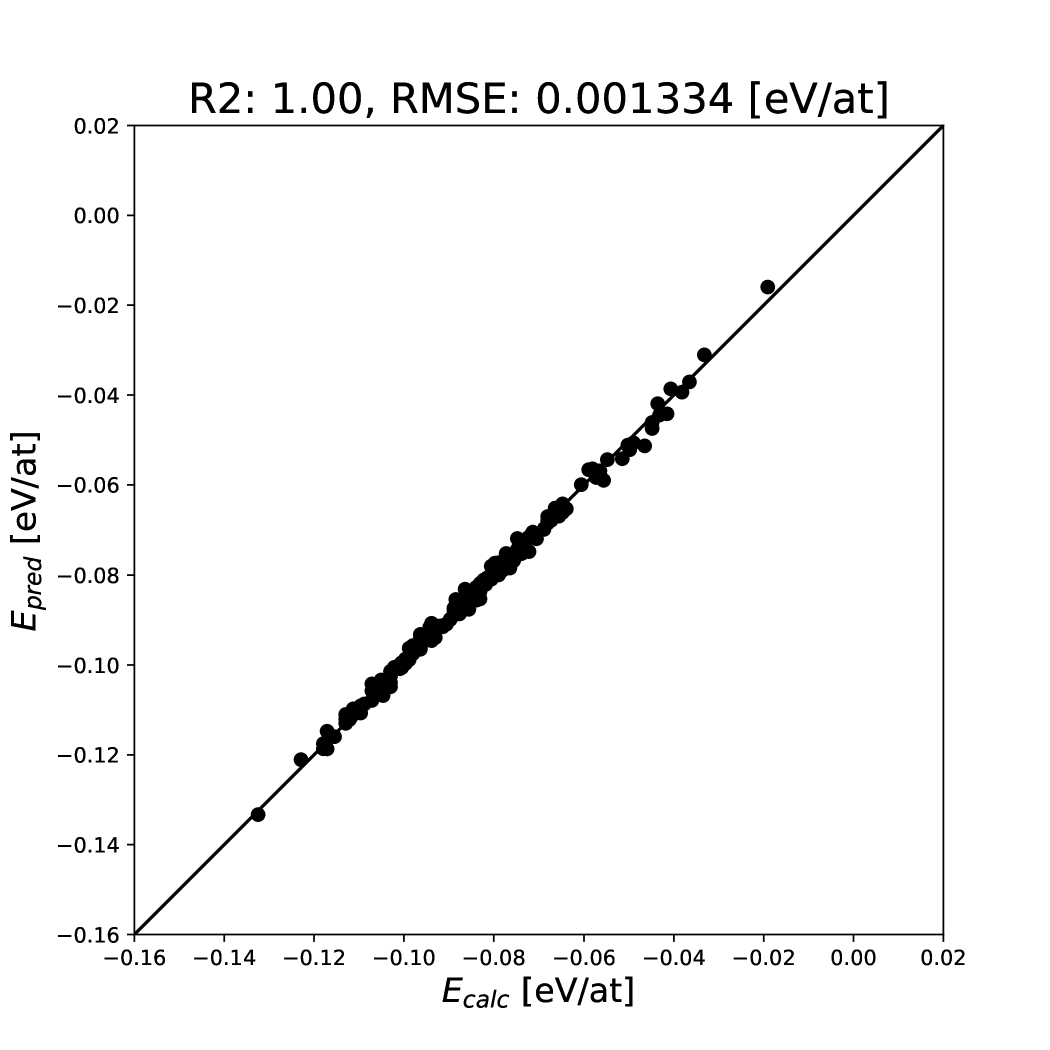}}\hfill
  \sublabel{c}{\includegraphics[width=.3\linewidth]{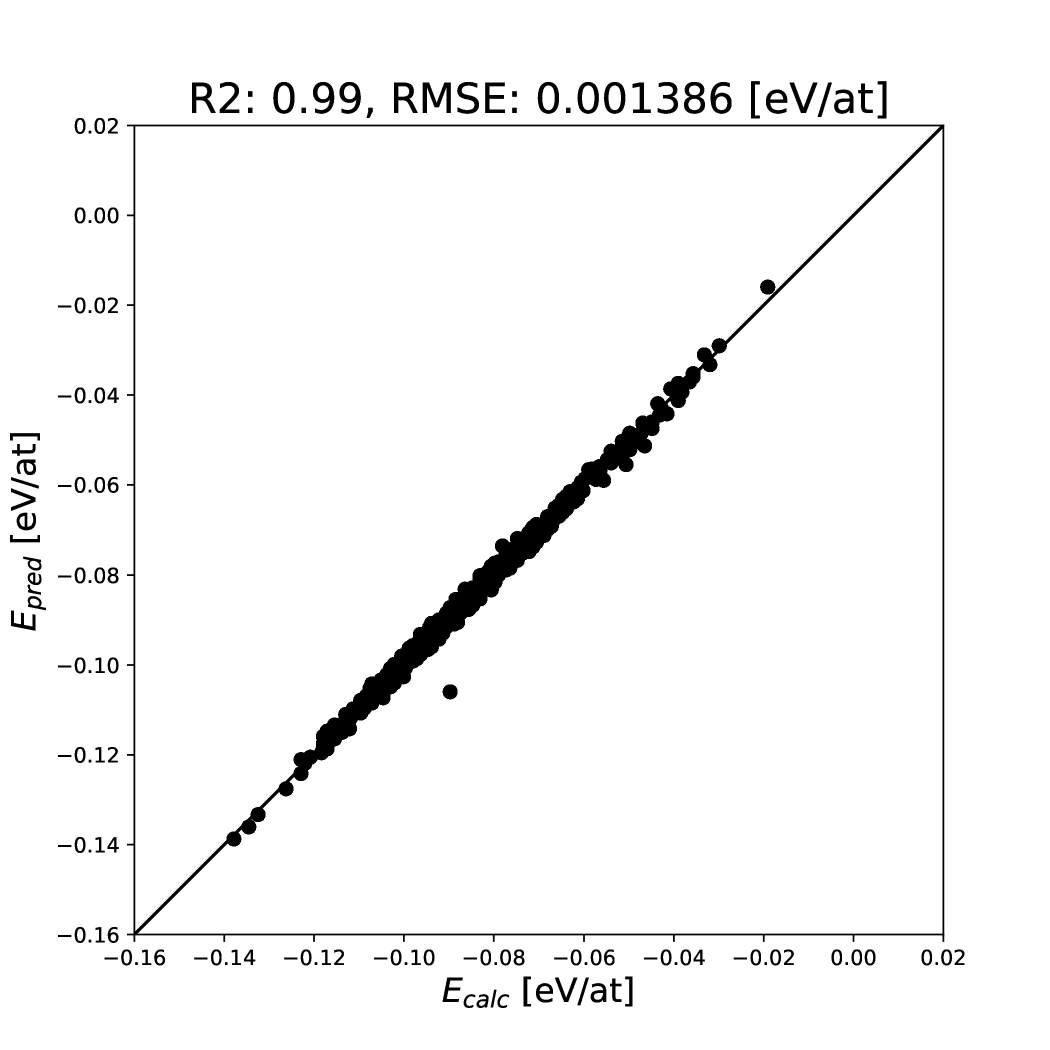}}
  \caption{The yyplot of the best model trained by 1247-point datasets for (a) training data, (b) validation data, and (c) test data.}
  \label{fig:yyplot_1247}
\end{figure*}

Figure~\ref{fig:yyplot_1247} shows the results of the best model trained by 1247-point datasets.
The test data RMSE is almost the same as the validation data RMSE, which indicates that the model is not overfitting and is trained appropriately.
The benchmark of this task RMSE is 0.001 eV/atom, which is the RMSE the simulations can achieve.
In the perspective of the benchmark, the RMSE of the test data is 0.00139 eV/atom, which is near the benchmark.
It means that our model can predict the energy of the crystal structure with almost the same precision as the simulations can achieve.

\begin{figure*}
  \centering
  \sublabel{a}{\includegraphics[width=.3\linewidth]{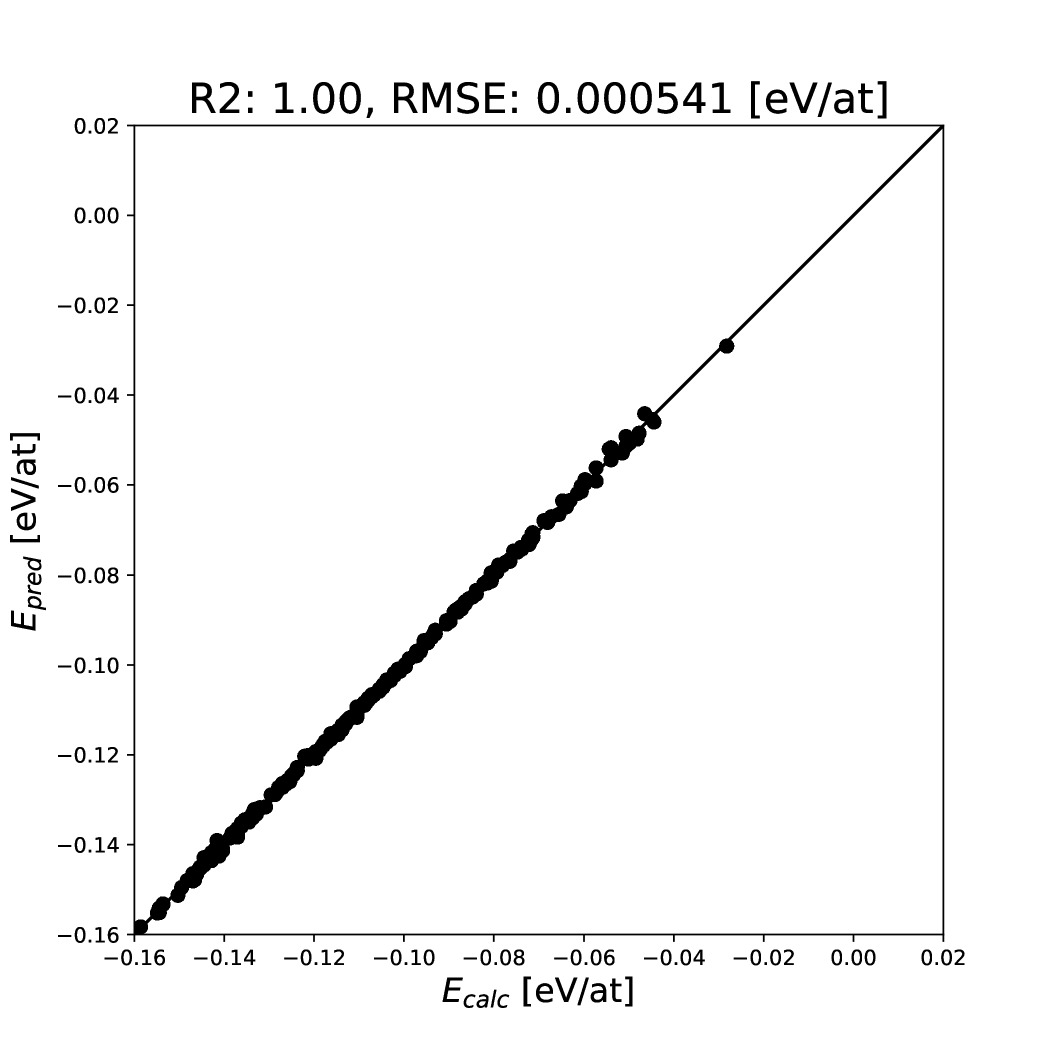}}\hfill
  \sublabel{b}{\includegraphics[width=.3\linewidth]{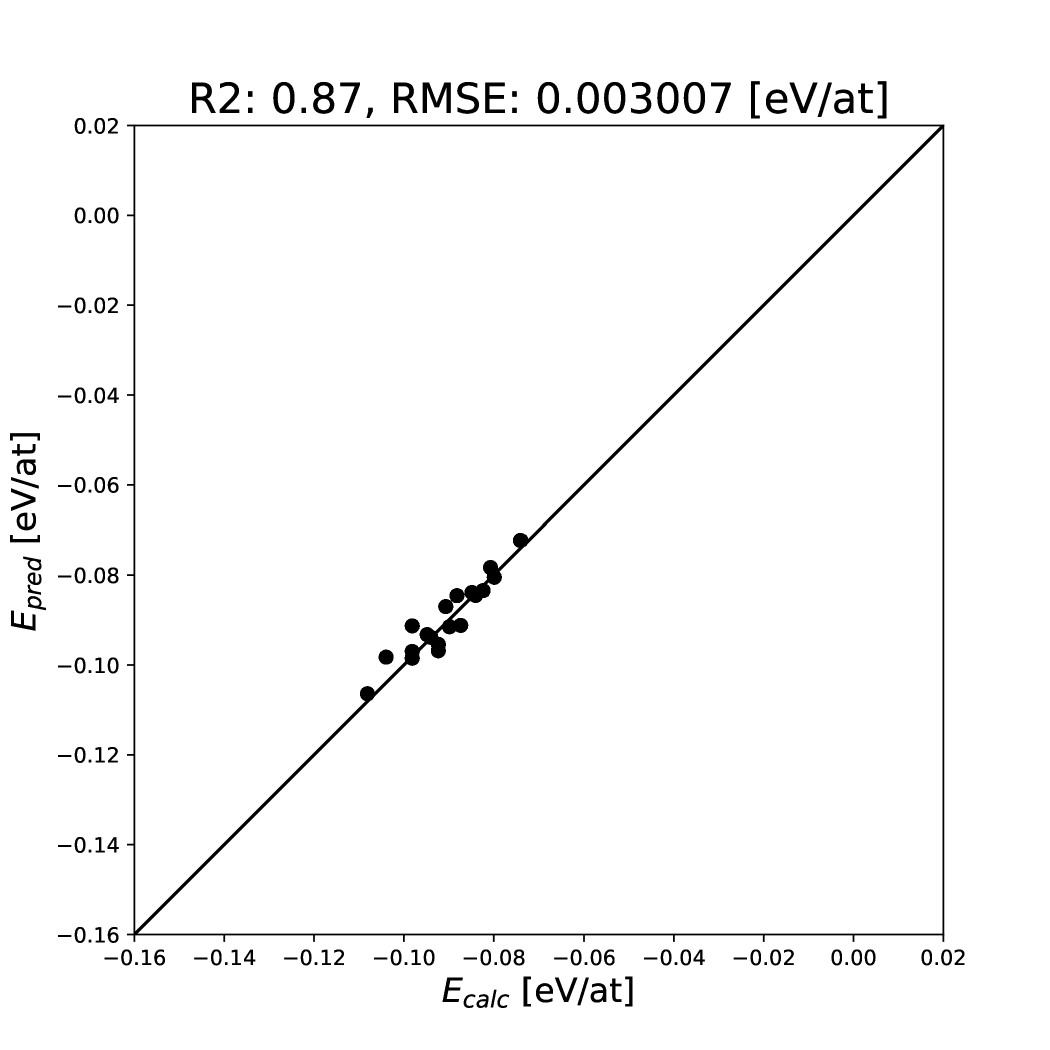}}\hfill
  \sublabel{c}{\includegraphics[width=.3\linewidth]{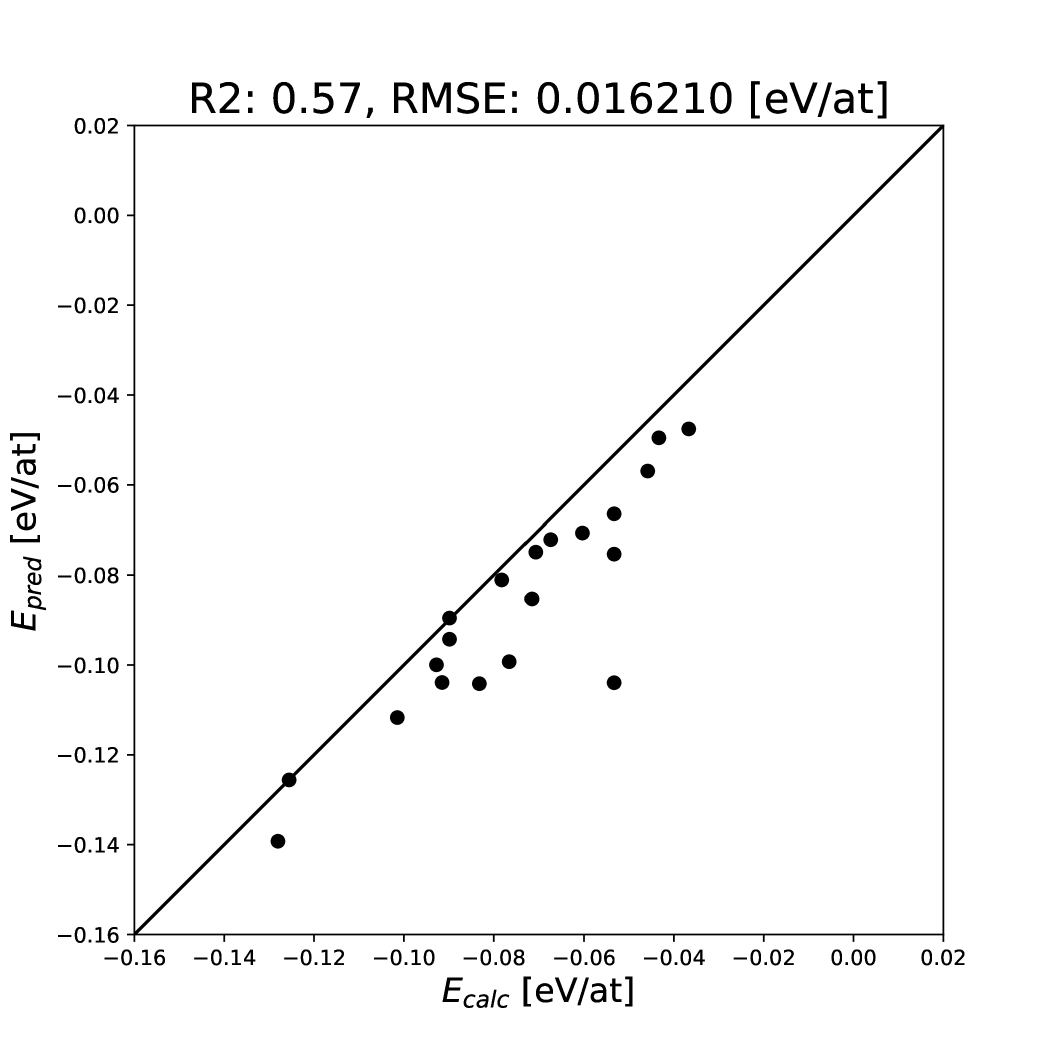}}
  \caption{The yyplot of the best model trained by 1247-point datasets for (a) low-energy test data, (b) 54-atom test data, and (c) 1024-atom test data.}
  \label{fig:yyplot_lowE}
\end{figure*}

Based on the above results, we evaluate the test data which have low energy.
Figure~\ref{fig:yyplot_lowE}(a) shows the low-energy test results of the best model trained by 1247-point datasets.
The RMSE of the low-energy test data is 0.00054 eV/atom.
This result indicates that our model can predict the energy of the crystal structure with high precision even when the energy of the crystal structure is low.

\begin{table}
    \centering
    \caption{RMSE of the training, validation, test and low energy test sets. Test set RMSE is almost same as benchmark RMSE and low energy test set RMSE archived greater performance than benchmark RMSE.}
    \label{RMSE_table}
    \begin{tabular}{c|ccccc}
        \hline
         & train & validation & test & low energy test \\
        \hline
        RMSE & 0.000968 & 0.00133 & 0.00139 & 0.00054 \\
        \hline
    \end{tabular}
\end{table}

\subsection{Scalability: tests of 54-atom and 1024-atom structures}

To evaluate the scalability of our model, we test the 54-atom and 1024-atom structures.
The capability to predict the energy of the crystal structure for 54-atom and 1024-atom structures is an important factor for practical use.
This is because the model is designed to predict the energy of the crystal structure with various sizes of atoms, and the model should be able to predict the energy of the crystal structure with high precision even if the size of the atoms is different.

Figure~\ref{fig:yyplot_lowE}(b) and (c) show the scalability test results of the best model trained by 598-point datasets.
The RMSE of the 54-atom test data is 0.00301 and the RMSE of the 1024-atom test data is 0.0162.
The RMSE of the 54-atom test data is approximately two times worse than the benchmark of the simulations and the RMSE of the 1024-atom test data is 15 times worse than the RMSE of the 1247-point datasets test data.
This result indicates that the model has difficulty in predicting the energy of the crystal structure with high precision when the number of atoms in the supercell differs from that used in training.
A likely cause is that the model was trained exclusively on 16-atom BCC supercells, so the local-environment features it has learned do not fully capture the long-range correlations and cumulative interactions that become increasingly important as the cell size grows.
In addition, because the total energy scales with the number of atoms, even small per-atom biases are amplified when the prediction is aggregated over a much larger supercell, resulting in substantially larger absolute errors.

\section{Discussion and Conclusion}
\label{sec:discussion}

The RMSE of the test data is almost the same as the RMSE of the validation data and achieves the benchmark RMSE.
These results show that our model is highly robust to symmetry-preserving transformations, specifically rotations and mirror operations, when predicting the energy of crystal structures.
In particular, the RMSE of the low-energy test data is lower than the benchmark RMSE, indicating that the model retains high precision in the low-energy regime, where accurate discrimination among energetically favorable structures is especially important for materials discovery and screening.
Taken together, these findings demonstrate that our machine-learning-based surrogate model can serve as a practical alternative to direct first-principles simulations for estimating crystal energies, enabling substantially faster evaluation of candidate structures under appropriate conditions.

However, the RMSE of the 54-atom and 1024-atom test data is considerably worse than the benchmark RMSE, and we observe a clear trend in which the prediction error grows as the number of atoms in the crystal structure increases.
This tendency indicates that the model is vulnerable to large crystal structures, likely because per-atom prediction errors accumulate over the many atoms of a larger supercell.
Moreover, the 1024-atom predictions tend to underestimate the actual energy, whereas the 54-atom predictions tend to overestimate it, implying that the model's prediction carries a systematic bias that becomes more pronounced as the crystal size increases.
Addressing this limitation is an important direction for future work.
In particular, identifying the origin of the bias and improving the model's ability to generalize across system sizes will be essential to extend its applicability to large-scale crystal structures and more realistic materials-design settings.

\section*{Acknowledgements}

This work was supported by the Future Scientist Exchange Program.

\bibliography{refs}

\end{document}